\documentclass[float,epsfig,usenatbib]{mn2e}

\usepackage{lscape}
\usepackage{graphicx}
\usepackage{appendix}
\usepackage{psfrag}
\usepackage{amsmath,amssymb}
\usepackage{threeparttable}
\usepackage{float}
\usepackage{subfigure}
\usepackage{afterpage}
\def\aaps{A\&AS}
\def\apj{ApJ}
\def\apjl{ApJL}
\def\mnras{MNRAS}
\def\pasp{PASP}
\def\aj{AJ}

\def\aap{A\&A}
\def\apjs{ApJS}
\def\nat{Nature}

\title[Diffuse FIR and FUV emission in Virgo]{Diffuse far-infrared and ultraviolet emission in the NGC4435/4438 
system: tidal stream or Galactic cirrus?}
\author[L. Cortese et al.]
{L. Cortese$^1$\thanks{luca.cortese@astro.cf.ac.uk}, G. J. Bendo$^2$, K. G. Isaak$^1$, J. I. Davies$^1$ \& B. R. Kent$^3$\\
  $^1$ School of Physics and Astronomy, Cardiff University, Cardiff CF24 3AA, UK. \\
  $^2$ Astrophysics Group, Imperial College London, Blackett Laboratory, Prince Consort Road, London SW7 2AZ, UK.\\
  $^3$ NRAO 520 Edgemont Road, Charlottesville, VA, 22903, USA.
}

\date{Accepted 2009 December 23.  Received 2009 December 16; in original form 2009 November 10}

\begin{document}
\newcommand{\Zsolar}{\mbox{$\,\rm Z_{\odot}$}}
\newcommand{\Msolar}{\mbox{$\,\rm M_{\odot}$}}
\newcommand{\Lsolar}{\mbox{$\,\rm L_{\odot}$}}
\newcommand{\xs}{$\chi^{2}$}
\newcommand{\dxs}{$\Delta\chi^{2}$}
\newcommand{\xsn}{$\chi^{2}_{\nu}$}
\newcommand{\ls}{{\tiny \( \stackrel{<}{\sim}\)}}
\newcommand{\gs}{{\tiny \( \stackrel{>}{\sim}\)}}
\newcommand{\asec}{$^{\prime\prime}$}
\newcommand{\amin}{$^{\prime}$}
\newcommand{\mstar}{\mbox{$M_{*}$}}
\newcommand{\hi}{H{\sc i}\ }
\newcommand{\hii}{H{\sc ii}\ }
\newcommand{\kms}{$\rm km~s^{-1}$}
\maketitle

\label{firstpage}

\begin{abstract}
We report the discovery of diffuse far-infrared and far-ultraviolet emission projected near the 
interacting pair NGC4435/4438, in the Virgo cluster.  This feature spatially coincides with 
a well known low surface-brightness optical plume, usually interpreted as tidal debris.
If extragalactic, this stream would represent not only one of the clearest examples of intracluster dust, 
but also a rare case of intracluster molecular hydrogen and large-scale intracluster star formation.  
However, the ultraviolet, far-infrared, \hi and CO emission as
well as the dynamics of this feature are extremely unusual for tidal
streams but are typical of Galactic cirrus clouds.
In support to the cirrus scenario, we show that a strong spatial correlation between far-infrared 
and far-ultraviolet cirrus emission is observed across the center of the Virgo cluster, 
over a scale of several degrees.
This study demonstrates how dramatic Galactic cirrus contamination can be, even at optical and ultraviolet 
wavelengths and at high galactic latitudes. 
If ignored, the presence of diffuse light scattered by Galactic dust clouds could significantly bias our 
interpretation of low surface-brightness features and diffuse light 
observed around galaxies and in clusters of galaxies. 
\end{abstract}

\begin{keywords}
	galaxies:individual: NGC4435/4438 -- galaxies: interactions -- ISM: dust.
\end{keywords}

\section{Introduction}
The study of the diffuse low surface-brightness light around galaxies and within 
clusters represents a unique tool for understanding the assembly history of 
structure in the universe. The detection of faint stellar tails associated 
with apparently undisturbed systems \citep{McConnachie09}, as well as the presence of 
diffuse light in galaxy clusters \citep{mihosvirgo}, are among the clearest 
pieces of evidence supporting a hierarchical picture of galaxy formation. 
In the past, the long integration times (and accurate flat-fielding) required to detect 
faint features have hampered the search for diffuse extragalactic light.  
In the next few years, with the advent of dedicated deep optical photometric surveys 
such as Pan-Starrs and the Next Generation 
Virgo Cluster Survey, will we eventually obtain an unprecedented view of the complex 
low surface-brightness universe, reaching very high sensitivities ($\mu$\gs28 mag arcsec$^{-2}$).
The main challenge for future investigations will likely be discriminating 
between extragalactic emission and scattered light from Galactic cirrus. 
At low surface-brightness ($\mu_{B}$\gs 27 mag arcsec$^{-2}$), scattered light from cirrus dust clouds is 
clearly detected at optical wavelengths even at high galactic latitude (e.g., \citealp{Guhathakurta89}). 
This might be erroneously interpreted as tidal debris or intracluster light.
To show how difficult it is to discriminate between extragalactic and Galactic low surface
brightness features, here we present the properties of a plume
projected near the NGC4435/4438 pair in the Virgo cluster. This plume
is usually interpreted as a tidal stream \citep{mihosvirgo}. 

NGC4438 (Arp 120) is the most dramatic example of a tidally disturbed galaxy in the Virgo cluster.
Dynamical models suggest that NGC4438 has been perturbed by a high-velocity ($\sim$800 \kms) 
interaction with the companion S0 NGC4435 (e.g., \citealp{vollmer05}). However, recent H$\alpha$ observations 
favour a different scenario in which NGC4438 has collided with the giant elliptical galaxy M86 \citep{kenney08}.  
Deep optical investigations \citep{mihosvirgo} also reveal the presence of a faint stellar plume apparently 
extending from NGC4435. This feature has been interpreted in the past as a tidal debris, but it 
is not clear how it would fit into the NGC4438-M86 interaction scenario.
In this Letter, we show that the multiwavelength properties of the optical plume are more similar to Galactic cirrus clouds 
than to tidal debris and that this feature might not be associated with the NGC4435/4438 system at all.

\begin{figure*}
\centering
\includegraphics[width=17.8cm]{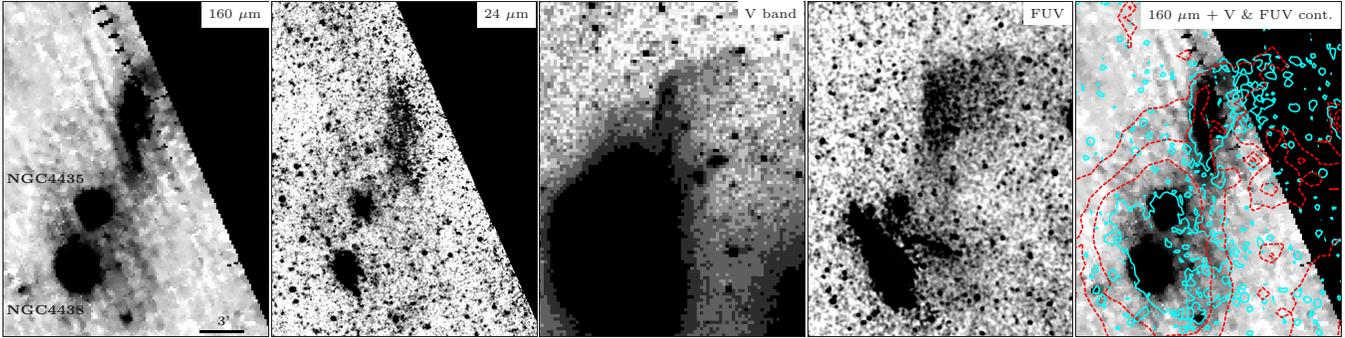}
\caption{\label{multistream} From left to right, the Spitzer MIPS 160$\mu m$, 24$\mu m$, V-band (Mihos et al. 2005) and GALEX-FUV images 
are shown. In the right panel, the optical (dashed red) and FUV (solid cyan) contours are superposed to the 
160$\mu$m. The contour levels are 25.5, 26.5, 27.5 mag arcsec$^{-2}$ and 29.8 AB mag arcsec$^{-2}$ in V and FUV, respectively. 
The 24$\mu$m and FUV images have been smoothed using a gaussian filter with $\sigma=$3 pixels in order to make the plume more visible.}
\end{figure*} 

\section{The data}
{\bf Spitzer observations.} 
We obtained Spitzer MIPS 24, 70 and 160 $\mu$m observations of the NGC4435/4438 system 
from the Spitzer Science Archive. These data are all scan maps taken as part of a single 
proposal (P30945). The image frames were created using the MIPS Data Analysis Tools \citep{gordon05} version 
3.10 and using the same technique described in \cite{bendo09}.
The rms in the final maps are 0.037, 0.38 and 0.39 MJy sr$^{-1}$ at 24, 70 and 160 $\mu$m, respectively.
The full widths at half maximum (FWHM) of the PSFs are 6, 18 and 38 arcsec at 24, 70 and 160 $\mu$m, respectively.\\
\noindent
{\bf Ultraviolet imaging.} 
The Galaxy Evolution Explorer (GALEX) has observed the NGC4435/4438 system in 
both near- (NUV, $\lambda$ = 2316 \AA, $\Delta \lambda$ = 1069 \AA) and far-ultraviolet 
(FUV, $\lambda$ = 1539 \AA, $\Delta \lambda$ = 442 \AA) bands, as part of the Nearby Galaxy Survey.
Two different tiles include the two galaxies and the 
stellar tail. We downloaded the final intensity maps from 
the MAST archive (GR4/5) and coadded them, reaching a total exposure time of 7628.35 
and 2967.8 sec in the NUV and FUV bands, respectively.
The spatial resolution of the final images is $\sim$5\arcsec. Details about the 
GALEX pipeline can be found in \cite{morrissey07}.\\ 
\noindent
{\bf Single-dish 21 cm \hi line data.} 
Single-dish \hi observations of the NGC4435/4438 system were obtained as part 
of the Arecibo Legacy Fast ALFA (ALFALFA) survey \citep{alfaalfa05}.   
The dataset used here comes from the first ALFALFA data-release \citep{giovanelli07}, covering 
the center of the Virgo cluster. The total integration time is 48 sec beam$^{-1}$, providing an 
rms of $\sim$2.2 mJy beam$^{-1}$ at a velocity resolution of $\sim$10 \kms. 
The data are gridded into a data cube with 1\arcmin pixels and the spatial resolution 
is given by the size of the Arecibo 
beam, 3.3\arcmin$\times$3.8\arcmin.\\
\noindent 
{\bf Single-dish CO(2-1) observations.} 
To detect CO(2-1) associated with the stellar tail, we have carried out a pilot observation using 
the single-pixel A3 receiver (beam-width $\sim$22\arcsec) at the James Clerk Maxwell Telescope (JCMT\footnote{The James Clerk Maxwell 
Telescope is operated by The Joint Astronomy Centre on behalf of the Science and Technology Facilities 
Council of the United Kingdom, the Netherlands Organization for Scientific Research, and the 
National Research Council of Canada.}). 
We used the full bandwidth of $\sim$1.9 GHz ($-$500\ls $V_{\odot}$\ls1900 \kms), thus covering the entire velocity range between NGC4435 and 
NGC4438 ($\sim$800 \kms),  with a velocity resolution of $\sim$1.3 \kms. 
A simple ON-OFF observing technique was adopted. 
The ON position roughly corresponds to the peak of 160$\mu$m emission in the stream 
($\alpha_{J2000}$=12:27:30.2, $\delta_{J2000}$=+13:12:29), while the OFF 
position was accurately chosen in order to avoid any possible 
contamination from Galactic emission. 
Data were reduced using the Starlink data reduction packages (KAPPA and SMURF). 
Individual spectra were inspected and then coadded to produce a final spectrum with 
a total on-source integration time of 1347s. 
The rms on the final baseline-subtracted spectrum is $\sim$0.018\degr~ K (expressed 
in antenna temperature).

\section{Results}
In Fig. ~\ref{multistream}, we compare
the Spitzer 24 and 160 $\mu$m images, the GALEX FUV image, and the
deep V-band image obtained by \cite{mihosvirgo} for the NGC
4435/4438 system.
They clearly reveal the presence of far-infrared and far-ultraviolet diffuse emission 
associated with the low surface-brightness ($\mu_{V}\sim$26.5-28 mag arcsec$^{-2}$) optical plume to 
the north/north-west of NGC4435 originally discovered by \cite{malin1994}.
While the MIPS field of view only includes the brightest part of this feature (close to NGC4435), 
the larger area covered by GALEX allows us to trace the whole extent of the stream 
out to a distance of $\sim$13\arcmin~from NGC4435.
The plume has an unusual L-shaped (or triangular) morphology, with abrupt and well 
defined edges to the east and north. No resolved sources (e.g., star-forming knots) are visible in the plume.
 
Interestingly, we first noticed the presence of diffuse emission in the Spitzer images, where 
the intensity of this feature is particularly high: $\sim$1.5-5 MJy sr$^{-1}$ and 
$\sim$0.05-0.08 MJy sr$^{-1}$ at 160 and 24 $\mu$m, respectively. 
A detailed inspection of IRAS images reveals that 
the plume is marginally visible in the 100 $\mu$m image, 
although the lower spatial resolution of IRAS made it impossible to clearly associate the plume with the 
NGC4435/4438 system.
No emission is detected at 70 $\mu$m, implying a 3$\sigma$ upper limit of 
$\sim$1.14 MJy sr$^{-1}$. 
The GALEX data reveal that the plume is really evident in FUV 
($\mu_{FUV}\sim$29 AB mag arcsec$^{-2}$), whereas in NUV it is only very marginally 
detected at a surface-brightness level of $\mu_{NUV}\sim$29.2 AB mag arcsec$^{-2}$.
\begin{figure}
\centering
\includegraphics[width=7cm]{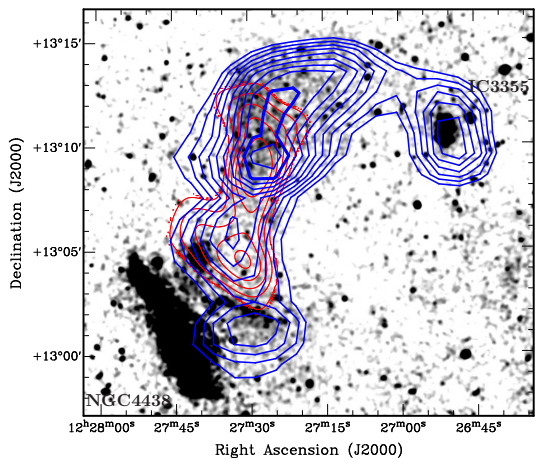}
\caption{\label{hi} The \hi contours of the plume superposed on the FUV image. The blue contours show the \hi emission 
detected by ALFALFA (including diffuse Galactic \hi) across the heliocentric velocity 
range $-$9.2$<V_{\odot}<-$19.5 \kms. Contours levels are 7.2 to 8.2 Jy \kms beam$^{-1}$, 
in steps of 0.1 Jy \kms beam$^{-1}$. As discussed in \S 3, the net \hi intensity of the plume has been obtained by 
performing a median subtraction of the Galactic emission around the feature.
For comparison, the VLA contours of the tail published by Hota et al. (2007) are shown in red.
We note that part of the \hi associated with NGC4438 and IC3355 is not visible given the narrow velocity 
range here investigated.}
\end{figure} 

\begin{figure}
\centering
\includegraphics[width=7cm]{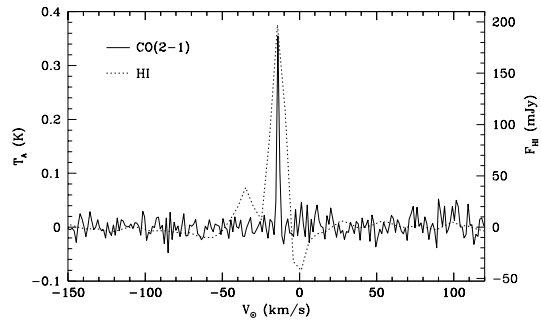}
\caption{\label{spectra} The single-pointing CO(2-1) (solid line) and integrated \hi (dotted line) spectrum 
of the plume.}
\end{figure}

In addition to dust and stars, this feature appears also to contain a significant amount of atomic hydrogen and CO. 
Using the Very Large Array (VLA), \cite{hota07} detected \hi emission associated with the 
brightest part of the plume. Intriguingly, the \hi has a low recessional velocity 
($\sim-$9.5 \kms, for a velocity resolution of 20.7 \kms) and a narrow velocity width ($<$40 \kms). 
Unfortunately, the northern and eastern edges of the tail were too close to the half-power of the 
VLA primary beam to be detected. Thanks to the higher sensitivity, velocity resolution and larger field of view 
of ALFALFA observations, we are now able to fill this gap and to characterize in more detail the 
\hi properties of this plume. By inspecting the cube through a velocity range -200$<V<$1000 \kms (i.e., the typical 
interval expected if the plume is related to the NGC4435/4438 system), we confirm that the only 
\hi emission clearly associated with the optical plume has a heliocentric velocity $V_{\odot}=-$15$\pm$4 \kms and 
a width $W_{50}=$5$\pm$7 \kms. 
In Fig.~\ref{hi} the ALFALFA and \cite{hota07} \hi contours are superposed to the 
FUV image of the NGC4435/4438 system. 
While the two datasets fairly agree in the overlapping region, the ALFALFA observations are able to trace the 
\hi along the whole extent of the plume, showing a good match between the \hi and FUV morphology of this feature.
To estimate its net \hi intensity we performed a median subtraction of the Galactic emission around the plume. 
The net flux density varies in the range 0.6-1.2 Jy \kms beam$^{-1}$. Assuming that the \hi fills 
the Arecibo beam, this can be translated into a column density of $\sim$1.5-3$\times$10$^{19}$ cm$^{-2}$ 
(see also \citealp{hota07}).
The integrated \hi spectrum of the plume\footnote{We only considered the main \hi feature visible Fig.\ref{hi}: i.e., the 
one clearly associated with the optical, UV and far-infrared plume.}  is shown in Fig.~\ref{spectra}. 
The peak and absorption features on either side of the detection are residual artifacts of the sky 
subtraction based on the median values in the selected channels. 
The recessional and velocity widths of the plume are confirmed, and even better constrained, by our CO(2-1) observations, 
as shown in Fig.~\ref{spectra}. CO(2-1) emission is clearly detected (peak signal-to-noise $\sim$20) at 
a heliocentric velocity $V_{\odot}=-$13.9$\pm$0.1 \kms and with a $FWHM=$1.5$\pm$0.2 \kms. 
 
In summary, the plume shows emission coming not only from stars and atomic hydrogen but also dust and CO, 
emerging as an almost unique feature in the panorama of low surface-brightness streams and tails 
discovered so far in the Virgo cluster.

\section{Discussion}
What is the nature of this plume? Is it really a tidal stream associated 
with the NGC4435/4438 system?
If extragalactic, it would represent an extremely rare example of 
intracluster dust and molecular hydrogen stripped by gravitational interactions. 
However, the multiwavelength properties of this feature demand a little bit of attention 
before one can interpret it as a tidal feature.
This is particularly true since the plume is detected in far-infrared, where the 
contamination from foreground Galactic cirrus emission might be significant, if not dominant.
In the following, we investigate the two possible scenarios (tidal stream and Galactic cirrus) in order 
to unveil the most likely origin of this feature.
\begin{figure*}
\centering
\includegraphics[width=8.8cm]{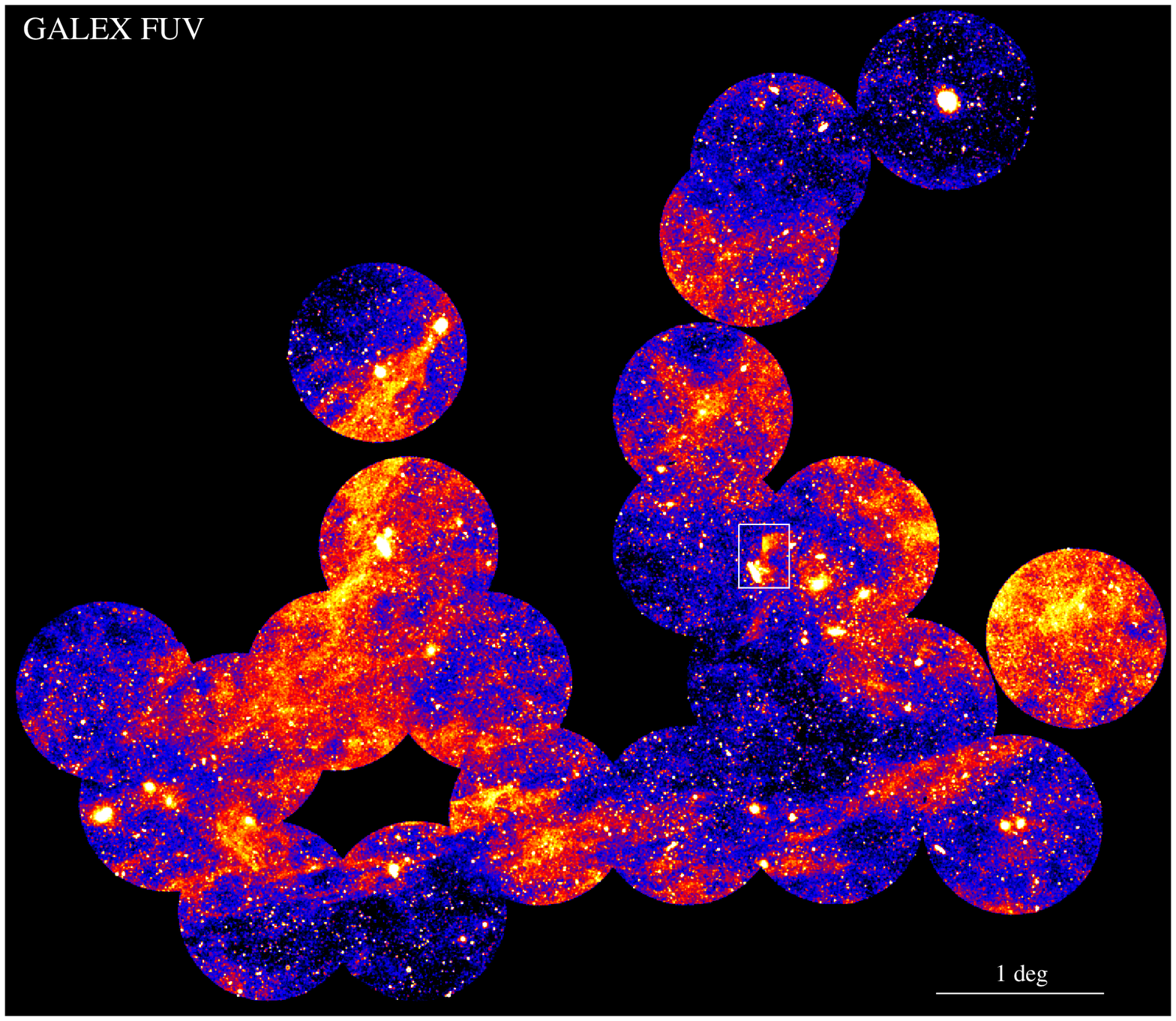}
\includegraphics[width=8.8cm]{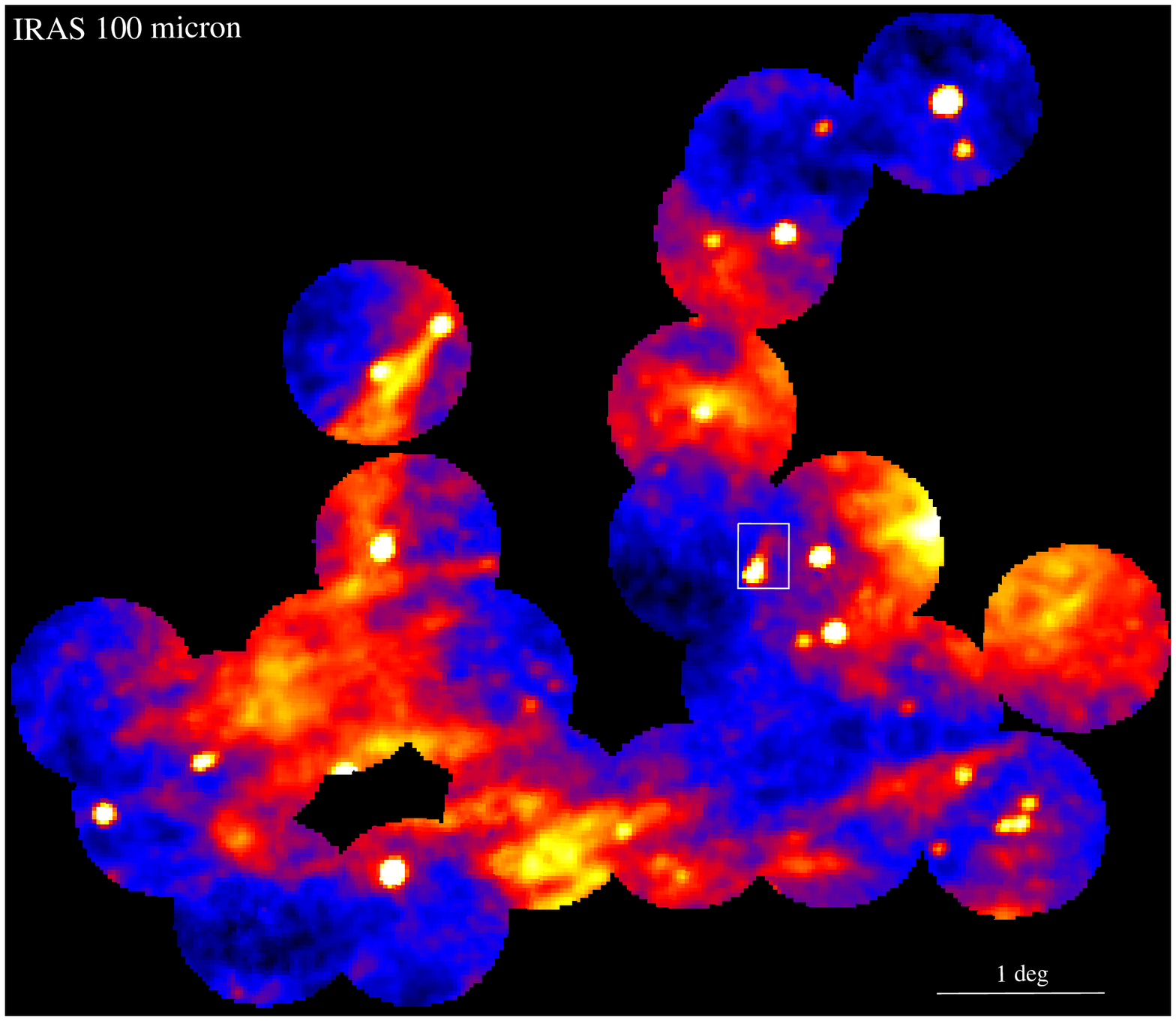}
\caption{\label{UVcirrus} The Virgo cluster region as seen in FUV (left) and 100 $\mu$m (right). 
The region shown in Fig.~\ref{multistream} is indicated by the white rectangle. The FUV mosaic have been smoothed using a 
gaussian filter with $\sigma=$9 pixels in order to highlight the diffuse cirrus emission.}
\end{figure*} 
 
\subsection{The tidal scenario}
If at the distance of the Virgo cluster (16.5 Mpc, \citealp{mei07}), the plume would have a physical size of 
$\sim$33 kpc (in both right ascension and declination), extending up to $\sim$70 kpc from NGC4438.
Its \hi and CO recessional velocities ($\sim-$14 \kms) are roughly consistent with  the \hi velocity of NGC4438 
($V_{\odot}=$104$\pm$2 and $W_{50}=$244$\pm$5 \kms, \citealp{giovanelli07}) but significantly different 
from the redshift of NGC4435 ($V\sim$801 \kms). This apparently rules out a gravitational 
interaction between NGC4438 and NGC4435: given their large velocity difference, we 
should expect the stripping material to lie at a velocity intermediate between 
the two systems. 

The plume apparently connects NGC4438 to the dwarf irregular galaxy IC3355 
($V\sim-$17 \kms, \citealp{goldmine}), perhaps suggesting a recent interaction between the two systems.
This seems supported by the presence of a \hi tail associated with IC3355 \citep{chung09}. 
However, contrary to what expected in case of a fly-by encounter with NGC4438, IC3355's tail 
points to the west, thus in the opposite direction with respect to NGC4438. 
Moreover, it is still unclear whether or not IC3355 is really a member of the M86 cloud \citep{chung09}. 
Thus, although we cannot exclude that IC3355 is playing a role, the current data do not allow us to 
further investigate this hypothesis.

A more plausible scenario is that the plume has been created 
during the collision between NGC4438 and M86 ($V\sim-$244 \kms) as recently unveiled by \cite{kenney08}.
A complex structure of H$\alpha$ filaments connects the two galaxies and spans a velocity range ($-$240$<V<$+70 \kms) 
consistent with the value observed in the plume. 
Moreover, the morphology of our feature (i.e., well defined edges on the east with less clear 
boundaries on the west side) would be consistent with the direction of motion of NGC4438, from 
west to east.
The time-scale of the interaction is $\sim$100 Myr \citep{kenney08}, 
significantly shorter than the dust sputtering time-scale expected at the position of NGC4438 
($\sim$10$^{8}$ yr; \citealp{popescu00}). Thus, although extremely unusual, the presence of dust 
in the tail does not rule out the tidal scenario. 

However, not all the properties of the plume are entirely consistent with a gravitation interaction.  
Above all, its velocity width ($\sim$2-5 \kms) is significantly smaller than the typical values ($\geq$20 \kms) 
observed in tidal tails, candidate tidal-dwarf galaxies, isolated \hi clouds and ram pressure stripped material 
(e.g., \citealp{hibbard01,kent07,osterloo05}). 
Although we cannot exclude that the cloud is seen `face-on', we consider 
this to be unlikely. The presence of a clear velocity gradient in the H$\alpha$ filaments 
between M86 and NGC4438 \citep{kenney08} suggests in fact that the tidal acceleration is not parallel to the plane of sky. 
Moreover, it is difficult to imagine how a violent process such as a tidal interaction would 
be able to strip material along a well defined direction and to keep the velocity dispersion of the \hi cloud so small.
It is therefore unclear how tidal debris can maintain such small velocity spread on a scale of a few tens of kpc.

Equally unusual are the UV properties of the tail. The FUV emission is in fact diffuse and not patchy with star-forming 
knots as observed in star-forming disks, tidal tails and extended UV disks (e.g., \citealp{thilker07}). 
Extraplanar diffuse UV emission is often interpreted as being due to light scattered by diffuse dust, in particular 
when associated with the halos of starburst galaxies (like M82; \citealp{hoopes05}). 
Although the Spitzer images reveal that dust is surely present in the plume, the lack of a powerful central 
starburst and the large distance of the tail from any strong radiation source apparently exclude the scattering scenario.
Thus, if extragalactic, the FUV emission must come from young stars in the plume. 
The absence of star-forming knots and H$\alpha$ emission \citep{kenney08} implies an age between 
$\sim$10 Myr (i.e., the lifetime of OB associations emitting in H$\alpha$) and $\sim$25 Myr (i.e., the lifetime of B stars, supposed 
to be responsible for the diffuse UV light in galactic disks; \citealp{pellerin07}), significantly shorter 
than the time-scale of the interaction but consistent with the age of the UV tidal tail to 
the west of NGC4438 discovered by \cite{n4438}. This might suggest that the plume has just entered a post-starburst phase, 
consistent with the fact that its colour is significantly bluer than the observed colour of NGC4438 ($FUV-V\sim$5.2 mag; 
\citealp{atlas2006}).     
However, if this is really the case, it remains unclear to us why the plume stopped forming stars. 
The very low velocity dispersion, the large amount of \hi and the presence of molecular hydrogen 
should allow the cloud to continue forming stars. Moreover, the fact that CO is usually associated with 
dense star-forming regions, even in tidal tails and stripped clouds \citep{lisenfeld04}, makes the post-starburst 
scenario unlikely.

\subsection{The Galactic cirrus hypothesis}
A completely different possibility is that the plume is in reality a Galactic cirrus cloud of gas and dust just 
superposed by chance near NGC4435. 
In this case, the diffuse FUV and optical emission would come from light scattered by the dust grains in the cloud. 

Before testing the cirrus hypothesis in detail, it is worthwhile to examine the probability of a
cirrus cloud appearing superimposed in front of the Virgo cluster.
Despite its high Galactic latitude ($b\sim$75\degr), the center of 
Virgo lies in a region that is significantly contaminated by cirrus emission. A ring-like cirrus structure centered near 
M87 and extending over several degrees \citep{brosch99} is clearly visible in the IRAS 100$\mu$m images 
(see Fig.~\ref{UVcirrus}, right). Since scattered light is often associated with cirrus clouds (e.g.,
\citealp{Guhathakurta89}), we should expect to see the same ring at both ultraviolet and optical wavelengths, 
though only at very low surface-brightness ($\mu_{B}\sim$27-28 mag arcsec$^{-2}$).
While deep optical data of the whole Virgo region are not available, the deep and wide GALEX coverage of the Virgo 
cluster makes it possible to test this hypothesis. We thus combined all the available GALEX pointings 
near the center of Virgo having FUV exposure time longer than 800 sec. 
The mosaic is shown in the left panel of Fig.~\ref{UVcirrus}, and it likely 
represents the highest resolution large-scale map of FUV cirrus emission made to date.  
It clearly illustrates the presence of diffuse FUV emission tracing the distribution of far-infrared 
Galactic cirrus emission. In addition the high spatial resolution of GALEX allow us to see for the first time the 
detailed structure of the cirrus clouds. 
In order to quantify the correlation between FUV and FIR diffuse emission, we first removed all sources detected by 
Sextractor \citep{sex} and then smoothed the UV images to the resolution of the `point-source free' \cite{schlegel98} IRAS 
map ($\sim$6 arcmin) by applying a Gaussian filter. 
The Pearson correlation coefficient between the FIR and UV intensity in each pixel (142$\times$142 arcsec$^{2}$) is $\sim$0.82, 
consistent with the range of values obtained by previous analysis of cirrus clouds (e.g., \citealp{haikala95,sasseen96}).
Thus, there is a reasonably high probability of detecting diffuse scattered far-ultraviolet 
(and perhaps also optical) light associated with cirrus emission in the direction of the Virgo cluster.

More direct support to the cirrus hypothesis is provided by the properties of the plume. 
Its negative recessional velocity, narrow velocity width and \hi column density are completely consistent 
with what is observed in Galactic low-velocity clouds (e.g., \citealp{stanimirovich06}).
In addition, the presence of CO emission (and thus of molecular hydrogen) is commonly 
observed in cirrus clouds \citep{reach98}. 

Several studies have shown that there is a good 
correlation between the 100$\mu$m far-infrared intensity and the \hi column density of cirrus clouds: 
e.g., $I(100\mu m)/N(HI)\sim$0.5-3 MJy sr$^{-1}$ / 10$^{20}$ cm$^{2}$ \citep{boulanger96}. 
At 160$\mu$m, the far-infrared to \hi ratio of the plume is $\sim$10 MJy sr$^{-1}$ / 10$^{20}$ cm$^{2}$.
Assuming a typical 160 to 100 $\mu$m flux ratio $\sim$ 1.5-3 for cirrus clouds \citep{bot09}, our feature 
appears to have a significant far-infrared excess. This is however not inconsistent with the cirrus hypothesis since 
Galactic `far-infrared excess clouds' are not rare at high Galactic latitudes and, as observed in this case, they 
are usually associated with significant CO emission \citep{reach98}.
Thus, the far-infrared-to-\hi ratio does not allow us to conclusively discriminate between the two proposed scenarios. 
This is also because our knowledge of the gas and dust content of cirrus is mainly based on studies of physical scales of 
tens of arcminutes, and it is not yet clear whether we can extrapolate them to the physical scale of the plume. 


\section{Conclusion}
In this Letter, we have investigated the multiwavelength properties of the optical plume projected near the 
NGC4435/4438 system. 
It is very tempting to interpret this feature as tidal debris in the Virgo cluster, as this would represent an 
extraordinary case of intracluster dust and molecular hydrogen as well as large-scale 
intracluster star formation.   
However, our analysis reveals that the dynamics, morphology, gas and dust content of this stream are more consistent 
with what observed in Galactic cirrus clouds than in tidal features associated with 
cluster galaxies. 
Whatever its real origin, this stream is a fascinating object and it clearly highlights how difficult is 
to discriminate between Galactic clouds and extragalactic diffuse light. 
This will be a great challenge for both far-infrared Herschel surveys and deep optical investigations of the 
extragalactic sky. Only a detailed characterization of the FUV-to-far-infrared properties of cirrus 
at all physical scales might eventually allow us to disentangle between the two different scenarios.

\section*{Acknowledgments}
We thank the anonymous referee for useful comments which improved the clarity of this manuscript.
We thank Ananda Hota and Chris Mihos for providing us with an electronic version of their data, Iain Coulson for his help in the preparation of the JCMT observations and David Hogg for useful discussions.
LC is supported by the UK Science and Technology Facilities Council and KGI by the Research Councils UK. BRK is a Jansky Fellow at the NRAO.

\end{document}